\documentclass[twocolumn,eqsecnum,aps,showpacs]{revtex4}

\begin{document}   
\title{Central charge and renormalization in supersymmetric
theories with vortices}
\author{K. Shizuya}
\affiliation{Yukawa Institute for Theoretical Physics\\
Kyoto University,~Kyoto 606-8502,~Japan }

\begin{abstract} 

Some quantum features of vortices in supersymmetric theories in 1+2
dimensions are studied in a manifestly supersymmetric setting of the
superfield formalism.  A close examination
of the supercurrent that accommodates the central charge and
super-Poincare charges in a supermultiplet 
reveals that there is no genuine quantum anomaly in the supertrace
identity and in the supercharge algebra, with the
central-charge operator given by the bare Fayet-Iliopoulos term alone. 
The central charge and the vortex spectrum
undergo renormalization on taking the expectation value of the
central-charge operator.  It is shown that the vortex spectrum is
exactly determined at one loop while the spectrum of the
elementary excitations receives higher-order corrections.
\end{abstract}


\maketitle

\section{Introduction} 

There has recently been considerable interest in
topological-charge anomalies in supersymmetric theories.
In the presence of topologically stable excitations the supercharge
algebra gets centrally extended and for the topological
excitations that saturate the Bogomol'nyi-Prasad-Sommerfield (BPS)
bound~\cite{B} classically the spectrum is determined exactly through
the central charge~\cite{WO}.  Multiplet shortening for BPS-saturated
excitations generally reveals that saturation persists at the quantum
level~\cite{WO,W}. 
It is, however, another question whether the spectrum of topological
excitations acquires substantial quantum corrections. 
Actually it took years until it was clarified from the study of 
solitons~\cite{Schon,KR,H,IM,CM,U,RN,NSNR,GJ,SVV,LSV,RNW,GRNW} in some
simple supersymmetric theories that topological charges do acquire quantum
anomalies~\cite{SVV},
with the soliton spectrum modified accordingly.   
This is mainly because direct calculations of the soliton spectrum 
and topological charge are a delicate task that requires proper 
handling of supersymmetry. 
In calculating the soliton
mass, for example, one has to handle an ill-defined sum over zero-point
energies of quantum fluctuations, and  may put the system in a finite box 
for mode regularization, which then easily spoils supersymmetry via the
boundary condition~\cite{GRNW}.

An ideal tool for keeping supersymmetry manifest is the superfield
formalism~\cite{SS,WB}. 
It allows one to study various quantum aspects of supersymmetric theories  
in a systematic way, avoiding such subtle problems as
encountered in finite-box calculations or in component-field calculations 
with non-supersymmetric regularizations.
Calculations with superfields are sometimes much more
efficient~\cite{GSR} than component-field calculations.

Fujikawa and van Nieuwenhuizen~\cite{FN}
developed a superspace approach to the central-charge anomaly.
Subsequently a superfield formulation of the central-charge anomaly  
was put forth~\cite{KScc} by making use of a supercurrent~\cite{FZ,CPS}
that places the topological charge, supercharges, energy and momentum 
in a supermultiplet.

Use of the superfield supercurrent makes manifest the supermultiplet nature of
various symmetry currents, conservation laws and the associated anomalies. 
This in turn reveals that the problem of topological-charge anomalies is
essentially the problem of the supertrace of the supercurrent.
This point of view has been verified for solitons and domain
walls in some low-dimensional theories, revealing a dual
(bosonic/fermionic) character of the central-charge
anomalies~\cite{KScc}; for solitons the anomaly derives from the
superconformal anomaly while for domain walls it apparently comes 
from induced spin.

The purpose of this paper is to extend a similar superspace analysis to vortices 
in 1+2 dimensions~\cite{Schm,LM},
for which somewhat nontrivial BPS saturation and finite renormalization of the
spectrum have recently been reported (within the component-field
formalism)~\cite{GRNW,Vass,RNWvor}.
We consider a one-parameter family of conserved supercurrents, 
which leads to a supercharge algebra of the same form and which
accommodates the improved superconformal currents for three and four
dimensions. It turns out that there is no genuine quantum anomaly in
the supertrace of the supercurrent and in the supercharge algebra, with
the central-charge operator given by the bare Fayet-Iliopoulos term
alone. 
We discuss how the central charge and the vortex spectrum undergo
renormalization on taking the expectation value of the central-charge
operator.  It is shown that the vortex spectrum is
exactly determined at one loop while the spectrum of the
elementary excitations receives higher-order corrections.

In Sec.~II we review some basic features of $N=2$ supersymmetric
theories with classical vortices in three dimensions. 
In Sec.~III we introduce a superfield supercurrent, and study its
classical conservation laws and improvement. 
In Sec.~IV we examine the supertrace of the supercurrent at the
quantum level and 
show the absence of a genuine superconformal anomaly.
In Sec.~V we discuss renormalization of the central charge and its
consequences.  Section~VI is devoted to a summary and discussion.

\section{N=2 supersymmetry in three dimensions}

Let us consider a supersymmetric version of the Abelian Higgs model 
with $N=1$ supersymmetry in four dimensions~\cite{Fayet}, 
described by the superspace action
\begin{equation}
S = \int\! d^{6}z\, {1\over{2}}\, W^{\alpha} W_{\alpha} 
+\int\! d^{8}z \left[-2\kappa V + \bar{\Phi} e^{2eV}\Phi \right], 
\label{superHiggs}
\end{equation}
where $z=(x^{\mu}, \theta_{\alpha},
\bar{\theta}^{\dot{\alpha}})$ denotes the superspace
coordinate, $d^{8}z=d^{4}x d^{2}\theta  d^{2}\bar{\theta}$, $d^{6}z=d^{4}x
d^{2}\theta$ and $W_{\alpha} = -{1\over{4}}\,
\bar{D}^{2}D_{\alpha}V$.
[We adopt superspace notation of Ref.\onlinecite{WB} but with metric
$(+ - - -)$; 
$D_{\alpha}= \partial/\partial \theta^{\alpha} -
(\sigma^{\mu}\bar{\theta})_{\alpha} p_{\mu}$ and
$\bar{D}^{\dot{\alpha}}  = \partial/\partial\bar{\theta}_{\dot{\alpha}} 
- (\bar{\sigma}^{\mu}\theta)^{\dot{\alpha}}p_{\mu}$ 
with $p_{\mu}= i\partial_{\mu}$.]
The Fayet-Iliopoulos (FI) term~\cite{FIL}
$-2\kappa V$ serves to introduce spontaneous breaking of U(1) gauge
invariance.

We shall study vortex solutions~\cite{Schm,LM} in the Higgs model 
with $N=2$ supersymmetry in three dimensions, which is obtained 
from this four-dimensional (4d) $N=1$ model by dimensional reduction.
The 4d model, actually, is afflicted with a gauge
anomaly (because of chiral gauge coupling to
matter) but this poses no problem for the reduced 3d model.  
One may instead start with an anomaly-free SQED
(supersymmetric quantum electrodynamics~\cite{WZQED}) version of 
the model~\cite{DDT} with oppositely-charged matter fields, i.e.,
\begin{equation}
\bar{\Phi} e^{2eV}\Phi \rightarrow \bar{\Phi} e^{2eV}\Phi 
+\bar{\Phi}_{-} e^{-2eV}\Phi_{-}
\label{QEDversion}
\end{equation}
in Eq.~(\ref{superHiggs}).
We study both models but handle the single-$\Phi$ model for
exposition. 
In addition, for conciseness, we use 4d notation until the reduction
to three dimensions is really needed.

The matter supermultiplet consists of a charged scalar field $A(x)$ 
and a charged fermion $\psi_{\alpha}(x) = (\psi_{1},\psi_{2})$ 
along with an auxiliary field $F$, with the chiral superfield
\begin{equation}
\Phi (z) 
= A(y)  + \sqrt{2}\theta\, \psi (y)  + \theta^{2}\, F(y),
\end{equation}
where $y^{\mu}\equiv x^{\mu} - i\theta\sigma^{\mu}\bar{\theta}$,
$\theta\, \psi = \theta^{\alpha} \psi_{\alpha}$, etc.
The real superfield $V(z)= V_{\rm WZ}+\Phi_{V}+\bar{\Phi}_{V}$
contains a gauge field $a_{\mu}(x)$ and 
a gaugino field $\chi_{\alpha}(x) = (\chi_{1},\chi_{2})$, 
along with an auxiliary field $D(x)$,
\begin{equation}
V_{\rm WZ} = \theta \sigma^{\mu}\bar{\theta}\, a_{\mu}
+ \theta^{2}\bar{\theta}\bar{\chi}
+ \bar{\theta}^{2}\theta \chi
+ {1\over{2}}\theta^{2}\bar{\theta}^{2} D ,
\end{equation}
where $(\sigma^{\mu})_{\alpha\dot{\alpha}} 
= (1,\sigma^{k})_{\alpha\dot{\alpha}}$, $\bar{\theta}\bar{\chi}
= \bar{\theta}_{\dot{\alpha}}\bar{\chi}^{\dot{\alpha}}$, etc.
The action~(\ref{superHiggs}) is invariant under gauge transformations 
$V \rightarrow  V + i\Lambda  -i\bar{\Lambda}$,  
$\Phi \rightarrow   e^{-2ie\Lambda}\Phi$ and $\bar{\Phi} \rightarrow  
\bar{\Phi} e^{2ie\bar{\Lambda}}$ with an arbitrary chiral superfield
$\Lambda (z)$ with $\bar{D}_{\dot{\alpha}}\Lambda
=D_{\alpha}\bar{\Lambda} = 0$.

In taking the Wess-Zumino gauge one retains only $V_{\rm WZ}$ for $V$
and sets $\Phi e^{2e\Phi_{V}}\rightarrow \Phi$ anew.
Supertranslations
\begin{equation}
x'^{\mu}= x^{\mu} +i\xi \sigma^{\mu}\bar{\theta} -
i\theta\sigma^{\mu}\bar{\xi} , \ 
\theta'= \theta + \xi,\ \ \  \bar{\theta}' = \bar{\theta} 
+ \bar{\xi}, 
\label{supertransl}
\end{equation}
on  $V_{\rm WZ}$ then yield the supersymmetry
transformation laws of the component fields
\begin{eqnarray}
\delta a^{\mu} &=& \xi\sigma^{\mu}\bar{\chi}
-\bar{\xi}\bar{\sigma}^{\mu}\chi, \nonumber\\
\delta \chi_{\alpha} &=&  \xi_{\alpha} D
- {1\over{2}}\, (\sigma^{\mu\nu}\xi) f_{\mu\nu}  , \nonumber\\
\delta D &=& -i( \xi \sigma^{\mu}\partial_{\mu}\bar{\chi}
+ \bar{\xi}\bar{\sigma}^{\mu} \partial_{\mu}\chi),
\end{eqnarray}
where $f_{\mu\nu} = \partial_{\mu}a_{\nu} - \partial_{\nu}a_{\mu}$.
They at the same time entail  gauge transformations  
$i\delta \Lambda_{V} =\theta\sigma^{\mu}\bar{\xi}\, a_{\mu}(y) +
\theta^{2}\bar{\xi}\bar{\chi}(y)$, 
which combine with supertranslations on $\Phi$
to yield gauge-covariant supersymmetry transformations of the matter
fields,
\begin{eqnarray}
\delta A &=& \sqrt{2}\, \xi \psi, \nonumber\\
\delta \psi_{\alpha} &=& \sqrt{2}\,\xi_{\alpha} F
- i\sqrt{2}\, (\sigma^{\mu}\bar{\xi})_{\alpha}D_{\mu}A,  \nonumber\\
\delta F &=&  -i \sqrt{2}\, \bar{\xi}\bar{\sigma}^{\mu}
(D_{\mu}\psi) + 2e \bar{\xi}\bar{\chi}A,\ {\rm etc.},
\end{eqnarray}
with $D_{\mu} = \partial_{\mu} +i e a_{\mu}$.

In the Wess-Zumino gauge the action reads 
$S= \int d^{4}x\, {\cal L}$ with
\begin{eqnarray}
{\cal L}&=& - {1\over{4}}\, f_{\mu\nu}^{2}  
 + \chi {\slash\!\!\! p}\bar{\chi} +{1\over{2}}\,D^{2} 
- D (\kappa - e A^{*}A) + F^{*}F 
 \nonumber\\
&& + \bar{\psi} i\, \bar{\slash\!\!\!\! D} \psi
+ |D_{\mu}A|^{2}
-\sqrt{2} e\, (A\bar{\chi} \bar{\psi}
+ A^{*}\chi \psi),
\label{Scomponent}
\end{eqnarray}
where $\bar{\slash\!\!\! \! D} = \bar{\sigma}^{\mu} D_{\mu}$.
Eliminating the auxiliary field $D$ yields the potential 
$-(1/2)\, (\kappa - e A^{*}A)^{2}$.
For $\kappa/e >0$ the potential has a vanishing minimum 
$D=\kappa - e A A^{*} \rightarrow 0$ for 
$\langle A \rangle_{\rm vac} \equiv v = \sqrt{\kappa/e}$,
and  supersymmetry is kept exact while the gauge symmetry gets
spontaneously broken.  For $\kappa/e < 0$, in contrast, supersymmetry is
spontaneously broken while gauge invariance is kept manifest.

We now reduce the system~(\ref{Scomponent}) down to three
dimensions. For translation from Weyl spinors to 3d Dirac spinors 
we note the correspondence
$\bar{\psi}\bar{\sigma}^{\mu}\chi\leftrightarrow
\psi^{\dag}\gamma^{0}\gamma^{\mu}\chi$ or
$(\bar{\sigma}^{\mu})^{\dot{\alpha}\alpha}= (1, -\sigma^{k})
\leftrightarrow \gamma^{0}\gamma^{\mu}$.
It is convenient to take the Majorana representation for the Dirac
matrices  
\begin{equation}
\gamma^{0}= \sigma^{2}, \ \ 
\gamma^{1} = i\sigma^{3},\ \ 
\gamma^{3} = -i\sigma^{1}. 
\end{equation}
This yields
$\gamma^{2}=-\sigma^{2}\sigma^{2} =-1$, telling us to eliminate
the "2" axis;
we thus take $(x^{0}, x^{1}, x^{3})$ for the 3d coordinate.
We do not display the 3d form of the Lagrangian here,
and simply remark that on reduction $a_{2}(x)$ turns into a
(pseudo)scalar field with matter coupling 
$-ea_{2}\bar{\psi}\bar{\sigma}^{2}\psi$.

We take $\kappa > 0$ (with $e>0$), in which case the reduced 3d model
supports classical vortex solutions governed by the
Bogomolnyi equation~\cite{B}
\begin{equation}
(D_{3} \pm iD_{1})A =0,\ 
f_{13} \pm (e A^{*}A -\kappa)=0, 
a_{2}\!= 0.
\label{Beq}
\end{equation}

We take the upper sign for vortices of vorticity  $n= 1,2,\cdots$, 
with the boundary condition
$A(x) \rightarrow  v e^{in\theta}$ and $a_{k} \rightarrow - (n/e)
\partial_{k}\theta$ for $r\rightarrow \infty$, where $\theta = \arctan
(x^{1}/x^{3})$, while $A(x) \propto r^{n}$ and $a_{k}
\rightarrow 0$ for $r\rightarrow 0$.
A vortex of vorticity $n$ carries the flux $\int
d^{2}x\, f_{13} =  \oint dx^{k}a^{k} =2\pi n/e$ and 
has energy $E_{\rm cl}$  equal to the central charge
\begin{equation}
Z_{\rm cl} =\int d^{2}x\, \epsilon^{ij}\partial_{i}(\kappa a_{j} +
iA^{*}D_{j}A) =  2\pi n\, \kappa/e
\end{equation}
with $\epsilon^{13}=1$.

\section{Superfield supercurrent}

Our main task is to study possible quantum modifications
of the supercharge algebra in the presence of vortices,
using the superfield formalism. 
In this section, as the first step, we introduce an
appropriate supercurrent and study its properties 
at the classical level.

The supercurrent for the present model is readily guessed from 
that~\cite{WL,CPStwo} for supersymmetric quantum electrodynamics:
\begin{eqnarray}
{\cal R}_{\alpha\dot{\alpha}} &=& 
(\nabla_{\alpha}\Phi) (\bar{D}_{\dot{\alpha}}e^{2eV}\bar{\Phi}) 
-{w\over{2}}\, [D_{\alpha}, \bar{D}_{\dot{\alpha}}] (\bar{\Phi}
e^{2eV}\Phi)
\nonumber\\
&& -2  W_{\alpha}\bar{W}_{\dot{\alpha}}, 
\label{SCclass}
\end{eqnarray}
where $\nabla_{\alpha} = e^{-2eV}D_{\alpha}e^{2eV} 
=D_{\alpha} +2e (D_{\alpha}V)$ is the gauge-covariant spinor derivative
for $\Phi$. 
Here $w$ is a free (real) parameter. 
This gauge-invariant supercurrent
${\cal R}_{\alpha\dot{\alpha}}$ can also be derived from the
action $S$ directly by a superspace Noether theorem using 
(gauge-covariant) local superconformal transformations of the
form~\cite{KS}
\begin{eqnarray}
\delta V &=& -\Omega^{\alpha}W_{\alpha} 
- \bar{\Omega}_{\dot{\alpha}}\bar{W}^{\dot{\alpha}}, \nonumber\\
\delta \Phi &=& - {1\over{4}}\, \bar{D}^{2}
[\,\Omega^{\alpha}\nabla_{\alpha} 
+ {w\over{2}}\, (D^{\alpha}\Omega_{\alpha})] \Phi, 
\label{localsctransf}
\end{eqnarray}
where  $\Omega_{\alpha}(z)$ and $\bar{\Omega}_{\dot{\alpha}}(z)
= [\Omega_{\alpha}(z)]^{\dag}$ are general spinor superfields.
This shows that $w$ stands for the R-weight $w_{R}$ of $\Phi$, 
associated with the R transformation 
$\Phi (z) \rightarrow e^{-iw_{R}\alpha} \Phi (x, e^{i\alpha} \theta,
e^{-i\alpha} \bar{\theta})$. 
Normally the chirality constraint~\cite{CPS} on $\Phi$ fixes the R-weight,
i.e., $w_{R}= 2/3$ for four dimensions and 
$w_{R} = 1/2$ for three dimensions. 
In the present model, however, $w$ is left arbitrary owing to 
simultaneous U(1) invariance $\Phi \rightarrow e^{i\alpha}\Phi$ of
the action $S$.

The supercurrent ${\cal R}^{\mu}(z) = {1\over{2}}\, 
(\bar{\sigma}^{\mu})^{\dot{\alpha}\alpha}
{\cal R}_{\alpha\dot{\alpha}} (z)$ obeys the conservation law 
(without using equations of motion)
\begin{eqnarray}
2i\partial_{\mu}{\cal R}^{\mu}
&=& d_{\Phi} + ( W^{\alpha}D_{\alpha} 
- \bar{W}_{\dot{\alpha}}\bar{D}^{\dot{\alpha}}  )\,
\delta S/\delta V
\nonumber\\
&& + {w\over{2}} \Big[ D^{2}(\Phi\, \delta S/\delta \Phi)  
-\bar{D}^{2} (\bar{\Phi}\, \delta S/\delta \bar{\Phi})  
\Big],\nonumber\\ d_{\Phi} 
&=& - (De^{2eV}\Phi)(De^{-2eV} \delta S/\delta \Phi) \nonumber\\
&& + (\bar{D}e^{2eV}\bar{\Phi}) 
(\bar{D}e^{-2eV} \delta S/\delta \bar{\Phi}). 
\label{dSC}
\end{eqnarray}
Here 
\begin{eqnarray}
\delta S/\delta V &=& -2\kappa +2e\, \bar{\Phi} e^{2eV}\Phi 
- D^{\alpha}W_{\alpha},  \nonumber\\  
\delta S/\delta \Phi 
&=&- (1/4)\, \bar{D}^{2}  e^{2eV}\bar{\Phi},\ {\rm etc.}, 
\end{eqnarray}
are identities implying the equations of motion.  
Thus the current ${\cal R}^{\mu}$ is conserved classically, 
$\partial_{\mu}{\cal R}^{\mu}=0$, owing to the equations of motion.

Similarly one can calculate the supertrace of the supercurrent 
or the supersymmetric trace identity, 
\begin{eqnarray}
\bar{D}^{\dot{\alpha}}{\cal R}_{\alpha \dot{\alpha}}  
&=& {\cal O}_{\alpha} + 4 \kappa W_{\alpha}
+ {\cal A}_{\alpha} , 
\label{DVcl} 
\end{eqnarray}
and its hermitian conjugate $D^{\alpha}{\cal R}_{\alpha\dot{\alpha}}
=\bar{\cal O}_{\dot{\alpha}} + 4 \kappa
\bar{W}_{\dot{\alpha}} + \bar{\cal A}_{\dot{\alpha}}$,
with
\begin{eqnarray}
{\cal O}_{\alpha} &=& - {1\over{4}}\, (3w - 2)
\bar{D}^{2}D_{\alpha} (\bar{\Phi} e^{2eV} \Phi), \label{Oa} \\
{\cal A}_{\alpha} 
&=& 2 W_{\alpha} {\delta S\over{\delta V}} - 2(\nabla_{\alpha}\Phi)
{\delta S\over{\delta \Phi}}  
+ w  D_{\alpha} ( \Phi {\delta S\over{\delta \Phi}}). 
\label{Aalpha}
\end{eqnarray}

The right hand side of Eq.~(\ref{DVcl}) characterizes  explicit breaking
to (4d) superconformal symmetry. 
The  $4\kappa W_{\alpha}$ term comes from the FI term and ${\cal
A}_{\alpha}$, which vanishes classically,  is a potential candidate for
the superconformal anomaly.
In contrast, ${\cal O}_{\alpha}$ is not a genuine breaking term.
Indeed ${\cal O}_{\alpha}$ disappears for $w = 2/3$ while 
$\partial_{\mu}{\cal R}^{\mu} =0$ holds classically for arbitrary $w$.
This indicates that $w$ represents the freedom to
improve the supercurrent;
this is seen explicitly from Eq.~(\ref{Tsym}) below, where   
$w$ arises with a familiar improvement term~\cite{CCJ} in the
energy-momentum tensor.

The supercurrent ${\cal R}^{\mu}(z)$, expanded in components 
\begin{equation}
{\cal R}^{\mu}
= R^{\mu} - i\theta J^{\mu} + i\bar{\theta}\bar{J}^{\mu} 
- 2\theta\sigma_{\lambda}\bar{\theta}\, T^{\mu\lambda}
+ \cdots,
\end{equation}
properly contains the R current 
\begin{equation}
R^{\mu} = (w - 1) \bar{\psi}\bar{\sigma}^{\mu}\psi
+i w A^{*}\! \stackrel{\leftrightarrow}{D^{\mu}}\!\! A
+ \bar{\chi} \bar{\sigma}^{\mu} \chi,
\label{Rcurr}
\end{equation}
energy-momentum tensor $T^{\mu\lambda}$ and
supersymmetry currents $J^{\mu}_{\alpha}$ and 
$(\bar{J}^{\mu})^{\dot{\alpha}}$,
\begin{eqnarray}
J^{\mu}_{\alpha} &=& 
(1-w)\sqrt{2}\,[ i(\sigma^{\mu}\bar{\psi})_{\alpha} F 
+ (\sigma^{\nu}\bar{\sigma}^{\mu}\psi)_{\alpha} D^{*}_{\nu}A^{*}]
\nonumber\\  
&& -2ie w (\sigma^{\mu} \bar{\chi})_{\alpha} A^{*}A 
-\sqrt{2}\,w A^{*}\! \stackrel{\leftrightarrow}{D^{\mu}}\!\!\psi_{\alpha}
\nonumber\\ &&-i  \sigma_{\rho}\bar{\chi}\, 
( D g^{\mu\rho} - i f^{\mu\rho} - \tilde{f}^{\mu\rho}  )
\end{eqnarray}
with  $\tilde{f}^{\mu\rho} 
= {1\over{2}}\epsilon^{\mu\rho\alpha\beta}f_{\alpha\beta}$ 
and $\epsilon^{0123}=1$. 
Here $A^{*}\!\!\! \stackrel{\leftrightarrow}{D_{\mu}}\!\! \psi
\equiv A^{*}D_{\mu} \psi  - (D^{*}_{\mu}A^{*})\, \psi$ with 
$D^{*}_{\mu}=\partial_{\mu} +iea_{\mu}$, 
and the conjugate current $(\bar{J}^{\mu})^{\dot{\alpha}}$ 
is obtained from $J^{\mu}_{\alpha}$ by substitution 
$i (\sigma^{\mu}\bar{\psi})_{\alpha} 
\rightarrow i(\bar{\sigma}^{\mu}\psi)^{\dot{\alpha}}$,
$(\sigma^{\nu}\bar{\sigma}^{\mu}{\psi})_{\alpha} 
\rightarrow
(\bar{\sigma}^{\nu}\sigma^{\mu}\bar{\psi})^{\dot{\alpha}}$, etc.
Remember the presence of the "2" components such as ${\cal R}^{2}$ and  
$T^{\mu 2}$ in the 3d case as well.

The energy-momentum tensor $T^{\mu\lambda}$ is the one somewhat improved
from the canonical tensor, with the symmetric component
\begin{eqnarray} 
T^{\mu\lambda}_{\rm sym} 
&=&(D^{*\mu}A^{*})D^{\lambda}A
+(D^{*\lambda}A^{*})D^{\mu}A
-f^{\mu\alpha}f^{\lambda}_{\ \ \alpha}  
\nonumber\\
&& +{i\over{2}}\, (\bar{\psi}\bar{\sigma}^{\mu} 
{\stackrel{\leftrightarrow}{D^{\lambda}}} \psi)_{\rm sym} 
+ {i\over{2}}\, (\bar{\chi}\bar{\sigma}^{\mu}\!\!
\stackrel{\leftrightarrow}{\partial^{\lambda}}\!\! \chi)_{\rm sym}
\nonumber\\
&& - g^{\mu\lambda} \Big\{
|D_{\nu} A|^{2} - {1\over{4}}\, f_{\alpha\beta}f^{\alpha\beta} 
- {1\over{2}}\,  D^{2} + \triangle \Big\}
\nonumber\\ 
&& + {1\over{2}}\, w(g^{\mu\lambda}\partial^{2}
-\partial^{\mu}\partial^{\lambda})(A^{*}A), 
\label{Tsym}
\end{eqnarray}
where $2\triangle = (1- w)\triangle_{1} - w \triangle_{2}$ 
consists of ''equation-of-motion'' terms
\begin{eqnarray} 
\triangle_{1} &=& 
 \bar{\psi}_{\dot{\alpha}} 
{\delta S\over{\delta \bar{\psi}_{\dot{\alpha}}}} 
+ \psi^{\alpha} {\delta S\over{\delta \psi^{\alpha}}}  
+  F {\delta S\over{\delta F}}
+ F^{*} {\delta S\over{\delta F^{*}}},  \nonumber\\
\triangle_{2} &=& A^{*} (\delta S/ \delta A^{*})
+ A (\delta S/ \delta A).
\label{AnomProd}
\end{eqnarray}

The antisymmetric component
$T^{\mu\lambda}_{\rm asym} =  {1\over{2}}\, (T^{\mu\lambda}  
- T^{\lambda\mu})$ is given by
\begin{eqnarray}   
T^{\mu\lambda}_{\rm asym}
&=& -\epsilon^{\mu\lambda\nu\rho}\Big[ 
{1\over{4}}\,\partial_{\nu} \Sigma_{\rho}  
+ {1\over{2}}\, f_{\nu\rho} (eA^{*}A + D) 
\Big]
\nonumber\\
&&\ \ +\triangle^{\mu\lambda}_{\psi,\bar{\psi}} -
\triangle^{\mu\lambda}_{\chi,\bar{\chi}} , 
\label{Tasym}
\end{eqnarray}
with 
\begin{eqnarray}
\Sigma_{\rho} &=& (1 - 2w)\, \bar{\psi}\bar{\sigma}_{\rho}\psi
+ \bar{\chi}\bar{\sigma}_{\rho}\chi + 4(1- w)iA^{*}D_{\rho}A, \nonumber\\
\triangle^{\mu\lambda}_{\psi,\bar{\psi}} &=& (i/4)\, [
\bar{\psi}\bar{\sigma}^{\mu\lambda} \delta S/\delta \bar{\psi} 
+  \psi {\sigma}^{\mu\lambda} \delta S/ \delta \psi],  
\end{eqnarray}
where $\triangle^{\mu\lambda}_{\chi,\bar{\chi}}$ is defined by 
$\triangle^{\mu\lambda}_{\psi,\bar{\psi}}$ with $\psi\rightarrow \chi$.
Note that one can write
\begin{equation}
eA^{*}A + D = \kappa + \delta S/\delta D, 
\end{equation}
which implies that 
\begin{equation}
 T^{\mu\lambda}_{\rm asym}
=  - \epsilon^{\mu\lambda\nu\rho}\Big( {\kappa\over{2}}\, f_{\nu\rho}
+{1\over{4}}\,\partial_{\nu} \Sigma_{\rho}\Big) 
\label{classicalcc}
\end{equation}
classically. 
This in turn shows that the dimensionally-reduced 3d model
has nonzero central charge
\begin{equation}
\int d^{2}{\bf x}\, T^{02}_{\rm asym} 
= \kappa  \int d^{2}{\bf x}\, f_{13}  = 2\pi n \kappa/e
\label{Zclass}
\end{equation}
in the background of a classical vortex. 
Note here that $\oint dx^{k} \Sigma^{k} = 0$ 
since $\Sigma^{k} \rightarrow  0$  rapidly 
as ${\bf x} \rightarrow \infty$.

The supertrace~(\ref{DVcl}), equally applicable to the 3d and
4d cases, may be further reduced to a form suited for three
dimensions. Writing $\bar{D}^{\dot{\alpha}}{\cal R}_{\alpha\dot{\alpha}} 
= (\sigma_{\mu}\bar{D})_{\alpha}{\cal R}^{\mu}$
reveals that on reduction
$(\sigma_{2}\bar{D})_{\alpha}{\cal R}^{2}$ turns into a superconformal
breaking term.  Note that ${\cal R}^{2}$ may be rewritten as
\begin{eqnarray}
2{\cal R}^{2} &=&
(w -1)\bar{D}\bar{\sigma}^{2}D(\bar{\Phi} e^{2eV}\Phi) \nonumber\\
&& +\bar{D}_{\dot{\alpha}} (\bar{W}^{\dot{\alpha}} G)
+ G{\delta S\over{\delta V}}, 
\end{eqnarray}
where $G =\bar{D}\bar{\sigma}^{2}DV$.
One can thus write the 3d form of the supertrace as
\begin{equation}
(\sigma_{\mu'}\bar{D})_{\alpha}{\cal R}^{\mu'} 
= {\cal O}'_{\alpha} + 4\kappa W_{\alpha} 
+ {\cal A}_{\alpha} , \label{threedDV} 
\end{equation}
with ${\cal O}'_{\alpha} = {\cal O}_{\alpha} 
- (\sigma_{2}\bar{D})_{\alpha}{\cal R}^{2}$ cast in the form
\begin{eqnarray}
{\cal O}'_{\alpha} &=& 
 {1\over{4}}\, (1 - 2w)\bar{D}^{2}D_{\alpha}(\bar{\Phi}e^{2eV}\Phi)
\nonumber\\ 
&& + {1\over{4}}\, \bar{D}^{2}[(\sigma_{2}\bar{W})_{\alpha}G]
- {1\over{2}}\, \sigma_{2} \bar{D} \Big(G{\delta S\over{\delta V}}\Big).
\label{Oprime}
\end{eqnarray}

For $w=2/3$, ${\cal O}_{\alpha}=0$ in Eq.~(\ref{DVcl}) and 
the supercurrent with $w=2/3$ contains the improved superconformal
currents for four dimensions. Similarly, for $w=1/2$ the first
term in ${\cal O}'_{\alpha}$ above disappears  and this implies that
${\cal R}^{\mu}$ with $w=1/2$ contains the improved currents appropriate
for three dimensions. The second term in ${\cal O}'_{\alpha}$ derives
from the super-Maxwell term $W^{2}$ which breaks 3d conformal invariance
while  the third term, as an anomaly candidate, may better
be included in ${\cal A}_{\alpha}$.

Here special remarks are in order as to the combination 
$G =\bar{D}\bar{\sigma}^{2}DV=-D\sigma^{2}\bar{D}V \sim -2 a^{2}$, 
associated with the "reduced" dimension $x^{2}$.
It forms a linear supermultiplet $(D^{2}G = \bar{D}^{2}G=0)$, is gauge
invariant and is odd under parity.
Let us remember that parity in 1+2 dimensions~\cite{JT} corresponds to
inversions about one of the spatial axes,
$x^{1}$ or $x^{3}$. Under inversion $x^{1} \rightarrow 
x'^{1} = -x^{1}$, e.g., spinors undergo the change 
$\theta_{\alpha} =(\sigma^{12})_{\alpha}^{\ \beta}
\theta'_{\beta} =(\sigma^{3})_{\alpha}^{\ \beta}\, 
\theta'_{\beta}$, so that one has $\theta^{2} = -(\theta')^{2}$,
$\theta\sigma^{2}\bar{\theta} = - \theta'\sigma^{2}\bar{\theta}'$,
$\theta\sigma^{0}\bar{\theta} = \theta'\sigma^{0}\bar{\theta}'$, 
etc.
The component-field action~(\ref{Scomponent})  thereby tells us 
that $a_{0}$ and $D$  are even under parity while
$a_{2}$ and $A$ are odd.  Hence $V (z)$ 
is even under parity while 
$\Phi (x,\theta, \bar{\theta}) = - \Phi (x',\theta', \bar{\theta}')$
is odd, and the superfield action~(\ref{superHiggs}) is parity-invariant.
[The parity-violating Chern-Simons term is contained in the
combination $VD\sigma^{2}\bar{D}V$. A direct one-loop calculation 
(though we omit the details)
suggests that no Chern-Simons term is induced in the present 3d model.]

We have so far handled the single-$\Phi$ Higgs model. For its SQED
generalization~(\ref{QEDversion}), one can define 
the supercurrent ${\cal R}_{\alpha\dot{\alpha}}$ by 
including the oppositely-charged partner $\Phi_{-}$ as well, 
with the conservation laws modified in an obvious fashion.  
The model still supports the same classical vortex solutions~\cite{DDT}
governed by the Bogomolnyi equation~(\ref{Beq}) with $\Phi_{-} = 0$.

\section{Trace identities}

In this section we promote the supercurrent to the quantum level.
Let us first note that ${\cal A}_{\alpha}$ in the
supertrace identity~(\ref{Aalpha}) consists of products of the
form (fields)$\times$(equations of motion). Such operator products vanish
trivially  at the tree level. They, however, are potentially singular
and, when properly regulated, may not vanish at the quantum level,
leading to anomalies, as is familiar from Fujikawa's
method~\cite{F,Ftwo}.  
Accordingly we examine ${\cal A}_{\alpha}$ and determine the form of 
the supertrace identity at the quantum level; see Ref.~\cite{KS} 
for an early study of the superconformal anomaly along this line.

For actual calculations we employ the superspace 
background-field method~\cite{CW} and expand 
$V  = V_{\rm c} + U$ and $\Phi=\Phi_{\rm c} + \eta $ 
around the classical fields $V_{\rm c}$ and $\Phi_{\rm c}$.
The quantum fluctuations at the one-loop level, in particular,
are then  governed by the action quadratic in $(U, \eta,\bar{\eta})$,
\begin{eqnarray}
S_{\rm q}
&=&  \int d^{8}z\,  \Big[ \bar{\eta} \Xi \eta 
+2e (\bar{\Phi}_{\rm c} \Xi \eta +\bar{\eta}\Xi \Phi_{\rm c})U
+ {1\over{2}}\, U M U  \Big],
\nonumber\\
M &=& 2\, (-p^{2} + 2 e^{2} \bar{\Phi}_{\rm c}\Xi\Phi_{\rm c}), 
\label{Sq}
\end{eqnarray}
with $\Xi = e^{2eV_{\rm c}}$.
Here we have taken the superfield Feynman gauge by 
including the gauge-fixing term $-(1/8)(\bar{D}^{2}U)(D^{2}U)$ 
in $S_{\rm q}$; $-p^{2}$ in $M$ has been 
$(1/8)D^{\alpha}\bar{D}^{2}D_{\alpha}$ before this gauge fixing.
The Feynman gauge is suited for formal reasoning because
one can avoid inessential complications due to ghost fields
(although $U \eta$ mixing leads to a set of Feynman
rules not very efficient for practical calculations).

The original action $S$ as well as the one-loop action $S_{\rm q}$
clearly has background gauge invariance
\begin{eqnarray}
&&(\Phi_{\rm c}, \eta) \rightarrow  e^{-2ie\Lambda}(\Phi_{\rm c}, \eta),
(\bar{\Phi}_{\rm c}, \bar{\eta}) \rightarrow 
e^{2ie\bar{\Lambda}} (\bar{\Phi}_{\rm c}, \bar{\eta}), \nonumber\\
&&V_{\rm c} \rightarrow V_{\rm c} + i\Lambda - i \bar{\Lambda},\ \
U \rightarrow U, 
\label{bgdinv}   
\end{eqnarray}
with arbitrary chiral phases $\bar{D}_{\dot{\alpha}}\Lambda 
= D_{\alpha}\bar{\Lambda}=0$.
They also have invariance under parity.
The background gauge invariance and parity therefore are exact symmetries
of the effective action calculated from  $S$,
although both eventually get spontaneously broken in the physical vacuum
where $\langle \Phi \rangle_{\rm vac} \not = 0$. 
We shall regularize the theory so that both of them are kept manifest for
the effective action.  See Appendix A for regularization and Feynman
rules.  
Actually the anomaly ${\cal A}_{\alpha}$ of Eq.~(\ref{Aalpha}) is
somewhat modified via gauge fixing; see Appendix B for details.
Its exact form, however, is not needed for our discussion below,
which relies on the fact that the modified ${\cal A}_{\alpha}$ still
preserves both background gauge invariance and parity; our analysis thus
applies to any gauges as long as they respect such invariance.

The potentially anomalous product ${\cal A}_{\alpha}$, being of
short-distance origin, should consist of  local polynomials of 
$\Xi = e^{2eV_{\rm c}}$, $e\bar{\Phi}_{\rm c} \Xi$, 
and $e\Xi \Phi_{\rm c}$, which are  (i) background gauge invariant,
(ii) neutral under U(1), and  are (iii) spinor superfields of 
dimension 5/2 or less in units of mass and of order $\hbar^{0}$ or higher.
As for possible gauge-invariant combinations, one may think
of, e.g.,  
$e^{2}\bar{\Phi}_{\rm c}\Xi\Phi_{\rm c}$ of dimension 2, 
$e(W_{\rm c})_{\alpha} \equiv -(e/4) \bar{D}^{2}D_{\alpha}V_{\rm c}$ 
of dimension 3/2, and
$eD\sigma^{2}\bar{D}V_{\rm c}$ of dimension 1.

The structure of possible superconformal anomalies is not quite arbitrary
and is constrained owing to boson-fermion balancing~\cite{CPS}.  
One can form chiral or real superfields out of quantum corrections. 
If the anomalies form a chiral supermultiplet, 
their contribution to the supertrace identity 
$\bar{D}^{\dot{\alpha}}{\cal R}_{\alpha\dot{\alpha}}$ is of the form
${\cal A}_{\alpha} \sim D_{\alpha}$(chiral superfield), as seen explicitly
from the $ D_{\alpha} ( \Phi\, \delta S/\delta \Phi)$ term in
Eq.~(\ref{Aalpha}). This possibility, however, is excluded in the
present 3d case (where
$\bar{D}^{\dot{\alpha}}{\cal R}_{\alpha\dot{\alpha}}$
has dimension 5/2).  This is simply because 
there is no candidate of (scalar) chiral superfields of
dimension 2 or less.  (Note that $e^{2}W_{\rm c}W_{\rm c}$ and 
$e^{2}\bar{D}^{2}(\bar{\Phi}_{\rm c}\Xi\Phi_{\rm c})$, e.g., have
dimension 3.)

If, on the other hand, the anomalies are associated with a real
superfield, they form a linear supermultiplet
of the form $\bar{D}^{2}D_{\alpha}$(real superfield) 
in the 4d case~\cite{CPS}.
In the dimensionally-reduced 3d case, there is one more possibility:
Anomalies of the form $(\sigma_{2}\bar{D})_{\alpha}X$ are possible, as
seen from Eq.~(\ref{threedDV}), but such a superfield $X$ must be odd
under parity just like ${\cal R}^{2}$.
The simplest choice $X \sim V_{\rm c}$ therefore is excluded. 
Instead, a possible candidate of least dimension is 
$(\sigma_{2}\bar{D})_{\alpha}\bar{D}\bar{\sigma}^{2}D$(real superfield),
which is rewritten as 
$- (1/2) \bar{D}^{2}D_{\alpha}$(real superfield), 
thus reducing to the anomaly multiplet of the 4d case.

The search for a possible linear anomaly multiplet is simplified
if one notes that
$e^{2}\bar{D}^{2}D_{\alpha}(\bar{\Phi}_{\rm c}\Xi\Phi_{\rm c})$ has
dimension 7/2 so that it does not arise as a short-distance anomaly. 
This implies that ${\cal A}_{\alpha}$ is formed of $V_{\rm c}$ alone.
Then there are two candidates, $e(W_{\rm c})_{\alpha}$ of dimension
3/2 and $e\bar{D}^{2}D_{\alpha}(D\sigma^{2}\bar{D}V_{\rm c})$ of
dimension 5/2;  of these the latter is excluded because of a parity
mismatch.  Therefore the only possibility is 
\begin{equation}
{\cal A}_{\alpha} = c e W_{\alpha}
\label{anomalyall}
\end{equation}
in operator form, with a coefficient $c = c_{1} + e^{2}c_{2}$ necessarily
divergent at one loop and terminating at two-loop order.
Indeed, a direct one-loop calculation yields a linearly-divergent
coefficient 
$c_{1} = 1/(2\pi \sqrt{\pi \tau})$ (with a UV cutoff 
$\tau \rightarrow 0_{+}$), which derives from 
\begin{equation}
(\nabla_{\alpha}\Phi) {\delta S\over{\delta \Phi}} 
= {1\over{2}}\, c_{1}  e W_{\alpha} ; 
\label{Danomaly}
\end{equation}
see Appendix C for details.

Fortunately an anomaly of the form~(\ref{anomalyall}) is harmless and
is effectively removed from the supertrace identity~(\ref{DVcl}) by
redefining the bare FI coupling 
$\kappa \rightarrow \kappa' = \kappa -ec/4$, 
without causing any change in observables.  
In this sense there is no genuine anomaly, ${\cal A}_{\alpha} =0$ 
to all orders in $\hbar$.

Actually, the situation is much clearer for the SQED$_{3}$ version 
of the model.  
There the matter fields $(\Phi, \Phi_{-})$ are oppositely charged
and they give a vanishing contribution to the anomaly,
${\cal A}_{\alpha} =0$.
The supertrace identity thus takes the naive form for both models,
\begin{equation}
\bar{D}^{\dot{\alpha}}{\cal R}_{\alpha\dot{\alpha}}  
= {\cal O}_{\alpha}  + 4 \kappa W_{\alpha},
\label{DJquantum}
\end{equation}
along with its conjugate 
$D^{\alpha}{\cal R}_{\alpha\dot{\alpha}} 
= \bar{\cal O}_{\dot{\alpha}} + 4 \kappa \bar{W}_{\dot{\alpha}}$. 
Note here that ${\cal O}_{\alpha}$ may effectively be absorbed into 
$W_{\alpha} = -{1\over{4}}\, \bar{D}^{2}D_{\alpha}V$ on the right-hand
side by the replacement
\begin{equation}
V \rightarrow  V + \gamma\, \bar{\Phi} e^{2eV} \Phi.
\label{Veff}
\end{equation}
with $\gamma = (1/4\kappa)\, (3w -2)$.

The supertrace~(\ref{DJquantum}) implies that
$\bar{D}^{\dot{\alpha}}{\cal R}_{\alpha\dot{\alpha}}$ is a chiral
(spinor) superfield.
This property fixes the higher components in terms of $R^{\mu}$
and $J^{\mu}$ uniquely, 
\begin{eqnarray}
{\cal R}^{\mu}&=& R^{\mu} - i\theta J^{\mu} 
+ i\bar{\theta}\bar{J}^{\mu} 
- 2 \theta\sigma_{\lambda}\bar{\theta}\, T^{\mu\lambda} \nonumber\\
&& + {1\over{2}}\theta^{2}\bar{\theta} 
\bar{\slash \!\!\! \partial}J^{\mu} 
- {1\over{2}} \bar{\theta}^{2}\theta
{\slash\!\!\! \partial}\bar{J}^{\mu}  
+ {1\over{4}} \theta^{2}\bar{\theta}^{2}\partial^{2}R^{\mu},
\label{scfullcomp}
\end{eqnarray}
and at the same time yields the conservation law
\begin{equation}
\partial_{\lambda}T^{\mu\lambda} = 0.
\label{drhoT}
\end{equation}
 
Equation~(\ref{DJquantum}) accommodates superpartners of the
trace identity, 
which, for $w=2/3$, read  (in 4d notation)
\begin{eqnarray}
i (\sigma_{\mu}\bar{J}^{\mu})_{\alpha} &=&
 4\kappa \chi_{\alpha}, \\ 
T^{\mu}_{\mu}
&=& 2 \kappa D,  \\
 T^{\mu\lambda}_{\rm asym} &=& -
\epsilon^{\mu\lambda\nu\rho} \Big[ {\kappa\over{2}}\,
f_{\nu\rho} + {1\over{4}}\, \partial_{\nu}R_{\rho} \Big], 
\label{Tasymfromti} 
\end{eqnarray}
where we have used the formula 
$\epsilon^{\mu\nu\rho\tau}\sigma_{\rho\tau} = 2i \sigma^{\mu\nu}$.
The effect of ${\cal O}_{\alpha}$ is readily recovered if one
notes Eq.~(\ref{Veff}): 
The $\theta\sigma^{\mu}\bar{\theta}$ component of 
$\bar{\Phi} e^{2eV} \Phi$, in particular, is nothing but 
the gauge current
\begin{equation}
j_{\mu}^{\rm gauge} = \bar{\psi} \bar{\sigma}_{\mu}\psi 
+ iA^{*}\! \stackrel{\leftrightarrow}{D}_{\mu}\!\! A,
\end{equation}
and, on setting 
$a_{\mu} \rightarrow a_{\mu} - \gamma\, j_{\mu}^{\rm gauge}$
in $f_{\nu\rho}$ of Eq.~(\ref{Tasymfromti}), 
$T^{\mu\lambda}_{\rm asym}$ precisely agrees with the classical
expression~(\ref{classicalcc}).

Letting $D^{\alpha}$ act on Eq.~(\ref{DJquantum})
reveals that the supercurrent ${\cal R}^{\mu}$ is conserved 
at the quantum level
\begin{equation}
\partial_{\mu}{\cal R}^{\mu}= 0.
\end{equation}
This leads to a conserved-charge superfield 
(in obvious 3d notation)
\begin{equation}
\int\! d^{2}{\bf x}\, {\cal R}^{0}
= Q_{\rm R} - i\theta Q + i\bar{\theta}\bar{Q} 
- 2 \theta\sigma_{\mu'}\bar{\theta}\, P^{\mu'}
- 2 \theta\sigma_{2}\bar{\theta}\, Z,
\label{sfcharge}
\end{equation}
consisting of the R charge $Q_{\rm R}$, supercharges $(Q_{\alpha},
\bar{Q}^{\dot{\alpha}})$, three-momenta $P^{\mu'}= \int d^{2}{\bf x}\,
T^{0 \mu'}_{\rm sym}$ and  central charge $Z= \int d^{2}{\bf x}\,
T^{02}_{\rm asym}$, with other charges vanishing.
This charge superfield, upon supertranslations~(\ref{supertransl}), 
gives rise to the supercharge algebra 
\begin{equation}
\{Q_{\alpha}, \bar{Q}_{\dot{\alpha}}\} 
= 2 (\sigma_{\mu'})_{\alpha\dot{\alpha}}\, P^{\mu'} 
+ 2(\sigma_{2})_{\alpha\dot{\alpha}}\,Z ,
\label{superchargealgebra}
\end{equation}
and $\{Q_{\alpha}, Q_{\beta}\} = 0$.
Note here that, as seen from Eq.~(\ref{Tasymfromti}), 
$\int d^{2}{\bf x}\, T^{0k}_{\rm asym} =0$
for $k=1,3$ [since $a_{2} \rightarrow 0$ and $R_{2} \rightarrow 0$ 
for ${\bf x} \rightarrow \infty$] while the central charge 
$Z= \int d^{2}{\bf x}\, T^{02}_{\rm asym}$ is sensitive to the asymptotic
nontrivial winding of the vortex field $A(x)\sim v e^{in\theta}$,
\begin{equation}
Z = \kappa \oint dx^{k} a^{k}
= \kappa  \int d^{2}{\bf x}\, f_{13}. 
\label{Zquantum}
\end{equation}
As noted earlier, $\oint dx^{k} \Sigma^{k} = 0$ for $|{\bf x}|
\rightarrow \infty$ and $\Sigma_{\rho}$ does not contribute to $Z$.
There is a general way to conclude this.
One may imagine piercing the vacuum state with a
vortex adiabatically.
The flux is thereby detected by $\oint dx^{k} a^{k}$.
In contrast, $\Sigma_{\rho}$, being gauge-invariant,
is single-valued  and $\oint dx^{k} \Sigma^{k}$ stays vanishing.

The unique feature of the FI term $2\kappa V$ is that it is gauge variant
by itself, and, in view of Eq.~(\ref{Veff}), it is precisely this
gauge-variant portion of $V$ that detects the vortex flux.
This explains why the central charge $Z$ derives from the
bare FI term alone in Eq.~(\ref{Zquantum}).

In the present 3d models gauge invariance is kept exact (though
spontaneously broken).  This is seen from 
\begin{equation}
\Phi\delta S/\delta \Phi
= -  (1/4)\, \bar{D}^{2}(\bar{\Phi}e^{2eV}\Phi) = 0,
\label{susychiral}
\end{equation}
which implies exact conservation of the gauge current,
$\partial^{\mu}j_{\mu}^{\rm gauge} = 0$, and which follows from the
absence of (gauge-invariant) chiral superfields 
of dimension two.
In this connection, we remark that 
$\Phi\delta S/\delta \Phi = 0$ implies 
$\triangle_{1} + \triangle_{2}=0$ in Eq.~(\ref{AnomProd})
while Eq.~(\ref{Danomaly}) implies  
$\triangle^{\mu\lambda}_{\psi,\bar{\psi}} = {1\over{8}}\,  c_{1} e\, 
\epsilon^{\mu\lambda\nu\rho}f_{\nu\rho}$  in Eq.~(\ref{Tasym}).

\section{Renormalization}

In Eq.~(\ref{Zquantum}) we have seen that 
there is no genuine quantum anomaly in the central charge 
$Z= \kappa \oint dx^{k} a^{k}$.  It takes the same form as the
classical expression~(\ref{Zclass}). 
The key difference is that it is promoted to an operator,
and one now has to determine the physical central charge from the
expectation value 
$\langle Z\rangle = \langle {\rm vor}|Z |{\rm vor}\rangle$. 
In this section we discuss how the central charge undergoes
renormalization through this process.

As for renormalization, power counting reveals that 
only the FI coupling $\kappa$  requires infinite renormalization
for the single-$\Phi$ Higgs model.
In addition, a nonrenormalization theorem~\cite{FNPRS} tells us that the
quantum correction to the FI term arises only at the one-loop level.
It is illuminating to calculate such a quantum correction
over the vacuum (with spontaneous parity breaking 
$\langle \Phi \rangle =v$), 
using the Feynman rules  given in Appendix A. 
Combining
$e^{2eV_{\rm c}} \langle \eta \bar{\eta}\rangle$,
$2e^{2}v^{2}e^{2eV_{\rm c}} \langle UU\rangle$ 
and $2eve^{2eV_{\rm c}}(\langle U \eta\rangle 
+ \langle U \bar{\eta}\rangle)$ one obtains the one-loop
correction to  the FI term $-2\kappa V_{\rm c}$,
\begin{equation}
2e \langle x| { i\, e^{\tau p^{2}}\over{p^{2} 
- m^{2}}}|x\rangle\, V_{\rm c} 
= {e\over{2\pi}}\Big[ {1\over{\sqrt{\pi \tau}}} 
- m \Big]\, V_{\rm c},
\label{oneloopcorr}
\end{equation} 
where $\tau$ is an UV cutoff and $m = \sqrt{2}ev$. 
The divergent term $\propto 1/\sqrt{\tau}$ can be removed on setting 
$\kappa = \kappa_{\rm ren} + \delta \kappa$ and choosing the counterterm
$\delta \kappa = e/(4\pi \sqrt{\pi\tau})$, with the renormalized FI
coupling $\kappa_{\rm ren}$.

In the SQED$_{3}$ version of the model the divergent term is
precisely canceled by an analogous contribution from the
(massless) partner $\Phi_{-}$ and one is left only with the finite
correction $-(em/2\pi) \, V_{\rm c}$.  This SQED$_{3}$ model therefore is
a finite theory.

One might be tempted to regard the $-(em/2\pi)\, V_{\rm c}$ term
in both models as contributing to finite renormalization of the FI term,
which, however, is not quite right.  The meaning of this term becomes
clear if one notes that Eq.~(\ref{oneloopcorr}) should come 
from a {\em gauge-invariant} one-loop effective action of the form
\begin{equation}
\Gamma^{(1)}
= \int d^{7}z\, \Big[ 
{e\over{2\pi}} {1\over{\sqrt{\pi \tau}}} \, V_{\rm c} 
-{1\over{2\pi}}\, \sqrt{2e^{2}\bar{\Phi}_{\rm c}
\Phi_{\rm c}} e^{eV_{\rm c}}
\Big],
\label{Gammaone}
\end{equation}
where $d^{7}z = d^{3}x\, d^{2}\theta\,  d^{2}\bar{\theta}$.
Indeed, a direct calculation yields precisely this form of the action 
if one retains only terms with no $D_{\alpha}$ or
$\bar{D}_{\dot{\alpha}}$  acting on $(\Phi_{\rm c}, V_{\rm c})$, i.e.,
the leading long-wavelength components; see Appendix~A.
The first term of Eq.~(\ref{Gammaone}) acts to renormalize the FI
coupling while the second term, being gauge-invariant [just like
$\Sigma_{\rho}$ of Eq.~(\ref{classicalcc}) ], does not contribute to the
central charge $Z$. It is therefore not legitimate to treat this
second term  or the $-(em/2\pi)\, V_{\rm c}$ term as finite
renormalization of  the FI term. 
It derives from long-wavelength quantum fluctuations and rather
contributes to a finite shift of energy: 
Minimizing the effective action with respect to
$V_{\rm c}$ within the asymptotic (vacuum) region yields
\begin{equation}
\kappa_{\rm ren} = e \{v^{2}- m/(4\pi) \},
\label{kren}
\end{equation}
which relates $v = \langle \Phi \rangle_{\rm vac}$ to the central charge
$\propto \kappa_{\rm ren}$.

Renormalization of the FI term takes place only at the one-loop level.
This means, in particular, that the one-loop results 
$\kappa = \kappa_{\rm ren} + \delta \kappa$  
with $\delta \kappa = e/(4\pi \sqrt{\pi\tau})$ for the single-$\Phi$ model
and $\kappa = \kappa_{\rm ren}$ for the SQED$_{3}$ version are actually
exact.
Thus the bare FI term is a finite operator
under renormalization, and the Heisenberg operator
$\kappa V$ equals $\kappa_{\rm ren} V_{\rm in}$ 
in the interaction picture.
As a result, the central charge, upon taking the expectation
value $\langle Z\rangle  = \langle {\rm vor}|Z |{\rm vor}\rangle$,
undergoes renormalization 
\begin{equation}
\langle Z \rangle 
= \kappa_{\rm ren} \oint dx^{k} \langle a^{k}_{\rm in}\rangle 
= 2\pi n\kappa_{\rm ren}/e.
\label{Zexp}
\end{equation}
This, together with $\kappa_{\rm ren}$ of Eq.~(\ref{kren}), confirms an
earlier result~\cite{RNWvor,GRNW} and reveals that this apparently
one-loop result~(\ref{Zexp}) is actually exact to all orders in
perturbation theory. Remember, however, that Eq.~(\ref{kren}) which
relates $v= \langle \Phi \rangle_{\rm vac}$ to $\kappa_{\rm ren}$,
in general, has higher-order corrections [in powers of
$e^{3}/\kappa_{\rm ren} \sim O(\hbar)$].

Having determined the renormalized central charge, let us finally
look into some consequence of the supercharge
algebra~(\ref{superchargealgebra}).   
In terms of $Q_{\pm} = (Q_{1} \pm iQ_{2})/2$ and 
$Q^{\dag}_{\pm} = (\bar{Q}_{1} \mp i\bar{Q}_{2})/2$, it reads
\begin{equation}
\{Q_{\pm},  Q^{\dag}_{\pm}\}
= P^{0} \pm Z,\ 
\{Q_{-}, Q^{\dag}_{+} \} = -P^{3} + iP^{1}.
\end{equation}
The static vortex solution~(\ref{Beq}) is invariant under
supertranslations caused by $Q_{-}$, realizing BPS saturation 
$Q_{-}|{\rm vor} \rangle =0$ classically. 
The saturation persists at the quantum level.   
This is because the vortex belongs to a short multiplet,
realizing only half of the original $N=2$ supersymmetry. 
Then $Q_{-}|{\rm vor} \rangle =0$ implies that the vortex
spectrum is determined by the central charge or $\kappa_{\rm ren}$
exactly, 
\begin{equation}
\langle {\rm vor}| P^{0} |{\rm vor}\rangle  
=\langle {\rm vor}| Z |{\rm vor}\rangle 
= 2\pi n\kappa_{\rm ren}/e.
\label{vortexspect}
\end{equation}
Remember here that formal reasoning based on the algebra is justified 
since we have verified that the supercharge algebra has no quantum
anomaly.  In contrast to the exact vortex spectrum, the spectra 
$m \sim \sqrt{2}ev \sim \sqrt{2e\kappa_{\rm ren}}$ of the
elementary excitations $\eta$ and $U$,
degenerate owing to supersymmetry, are determined only approximately 
through $v$, which, as in Eq.~(\ref{kren}), 
is affected by higher-order long-wavelength fluctuations.

\section{Summary and discussion}

In this paper we have studied some quantum aspects of vortices in the
supersymmetric Higgs model in 1+2 dimensions and its SQED$_{3}$
generalization, by a close examination of the supercurrent and associated
conservation laws within the superfield formalism. 
The central-charge operator $Z= \kappa \oint dx^{k} a^{k}$
turns out to derive from the bare FI term alone and measures 
the vortex flux.
It acquires no (genuine) quantum anomaly but undergoes
renormalization so that the observable central charge, 
given by the expectation value 
$\langle {\rm vor}|Z| {\rm vor} \rangle$, is an integral
multiple of the renormalized FI coupling $\times$ flux quantum,
$n\kappa_{\rm ren}2\pi/e$, which, via BPS saturation, equals the vortex
spectrum.   Renormalization of the FI coupling $\kappa$ takes place only
at the one-loop level and the vortex spectrum $n\kappa_{\rm ren}2\pi/e$ 
is exactly known.
In contrast, the spectrum $m \sim \sqrt{2} e v$ of the elementary
excitations $\eta$ and $U$ receives corrections from  
higher-order quantum fluctuations.

Our analysis makes heavy use of the superfield supercurrent
${\cal R}^{\mu}$, but the choice of such a conserved supercurrent is not
unique (because of the simultaneous presence of gauge invariance and
R-symmetry).  We have considered a family of supercurrents
parameterized by the R-weight $w$ of the matter superfield, and noted that
it leads to a supercharge algebra of the same form.  This $w$ represents
the freedom to improve the supercurrent. The improved superconformal
currents thereby emerge as components precisely when one assigns the
canonical R-weight, i.e., 
$w=1/2$ in 1+2 dimensions and $w=2/3$ in 1+3 dimensions.

Finally we wish to emphasize that use of superfields (+ supersymmetric
regularization) allows one not only to keep supersymmetry manifest in
actual calculations but also to substantiate formal reasoning based on
supersymmetry. With such firm control on supersymmetry, in particular,
persistence of BPS saturation at the quantum level is a concrete
consequence of the supercharge algebra once multiplet shortening for
BPS-saturated excitations is concluded algebraically.  
The excitation spectrum is thereby determined from the central-charge
operator, with the effect of  renormalization entering on taking the
expectation value. The superfield formalism can thus neatly avoid some
subtle problems encountered in component-field calculations.

\acknowledgments

This work was supported in part by a Grant-in-Aid for Scientific Research
from the Ministry of Education of Japan, Science and Culture (Grant No.
14540261).


\appendix

\section{Calculations}
In this appendix we outline some one-loop calculations in the text, 
based on the superfield action~(\ref{Sq}).
Let us first consider the simplified action (in 4d notation)
\begin{equation}
S_{\rm q} = \int d^{8}z\, \bar{\eta}\, \Xi\, \eta
+\int d^{6}z\, \eta\, j 
+ \int d^{6}\bar{z}\, \bar{\eta}\, \bar{j}
\end{equation}
with $\Xi = e^{2eV_{c}}$ and  source terms $j(z)$ and
$\bar{j}(z)$. 
We solve the equation of motion
$1_{+}\Xi \eta + \bar{j}=0$ in such a way that background gauge
invariance  is kept manifest:
\begin{equation}
\eta = - {\cal D}^{-1}\, 1_{-}\Xi_{\rm c}^{-1}\bar{j},
\label{eqeta}
\end{equation}
with ${\cal D} = 1_{-}\Xi^{-1}1_{+}\Xi$, 
$1_{+}= -{1\over{4}}\, D^{2}$ and 
$1_{-}= -{1\over{4}}\, \bar{D}^{2}$.
Note here that ${\cal D}$ acting on $1_{-}$ is invertible
so that 
\begin{equation}
{\cal D}^{-1}1_{-} = [\Pi ^{2} - e W_{\rm c}^{\alpha}\nabla_{\alpha} 
- (e/2)\, (DW_{\rm c})]^{-1} 1_{-},
\end{equation}
with  $\Pi_{\mu}= p_{\mu} - ea_{\mu}$.
Substituting Eq.~(\ref{eqeta}) back into $S_{\rm q}$ yields a stationary
action of the form $i\int d^{6}z d^{6}\bar{z}' \, j(z) 
\langle \eta \bar{\eta}\rangle^{(0)} \bar{j}(z')$, from which one can
read off the propagator
\begin{equation}
\langle \eta (z) \bar{\eta}(z') \rangle^{(0)}
= i \langle z|  {e^{\tau {\cal D}}\over{\cal D}}\,
1_{-}\Xi^{-1}1_{+} |z'\rangle, 
\label{etapropagator}
\end{equation}
where for regularization we have introduced an UV cutoff 
$e^{\tau {\cal D}}$ with $\tau \rightarrow 0_{+}$ which respects
background gauge invariance, parity and supersymmetry. 

Let us now include the remaining terms of Eq.~(\ref{Sq}). 
The effect of $\eta U$ coupling  is readily taken care of by the
replacement 
$j \rightarrow j + 2e1_{-}\bar{\Phi}_{\rm c}\Xi U$ and 
$\bar{j} \rightarrow \bar{j} + 2e1_{+}\Phi_{\rm c} \Xi U$.
This yields a $U^{2}$ term, which modifies the kinetic term $M$ of $U$.
Minimizing the action with respect to $U$ then yields a set of relevant
propagators.  We omit the details here and simply remark that the
$\eta$ propagator thereby acquires corrections,
$\langle \eta \bar{\eta} \rangle = 
\langle \eta \bar{\eta}\rangle^{(0)} + O(\bar{\Phi}_{\rm c} 
\Phi_{\rm c})$.

In Eq.~(\ref{oneloopcorr}) we make a one-loop calculation over the
ordinary vacuum (with  $\langle \Phi \rangle =v$), based on 
the zero{\it th}-order action
\begin{equation}
S_{\rm q} = \int d^{8}z\, \Big[ 
\bar{\eta} \eta + 2ev(\eta  + \bar{\eta})U
+{1\over{2}}\, U M U \Big]
\end{equation}
with $M= 2 (-p^{2} + m^{2})$ and $m^{2}= 2e^{2}v^{2}$.
In this case, the $O(U^{2})$ term mentioned above modifies 
the kinetic term $M$ into 
$M' = M - 2m^{2} \{1_{-},1_{+}\}/p^{2} =
-2 [\, p^{2}+ (m^{2}/p^{2})Y]$ with $Y = (1/8)D\bar{D}^{2}D$,
where we have used the identity
$\{1_{-},1_{+}\} = p^{2} +Y$. One can readily invert $M'$ by
noting $Y^{2}=-p^{2}Y$ and $Y1_{+}=Y1_{-}=0$.
Here we quote the Feynman rules constructed in this way:
\begin{eqnarray}
\langle \eta \bar{\eta} \rangle &=&  
\langle z | {i\over{p^{2}}} \, ( 1 - {m^{2}\over{p^{2}}})  
1_{-}1_{+} |z'\rangle, \nonumber\\
\langle UU \rangle &=& -  \langle z |  {i\over{2 p^{2}}} \, 
\Big[ 1 - {m^{2}\over{p^{2} (p^{2} - m^{2}) }}  Y \Big]\, |z'\rangle,
\nonumber\\
\langle U\eta  \rangle &=& 
ev\, \langle z | {i\over{(p^{2})^{2}}} \, 1_{+}1_{-} |z'\rangle,
\nonumber\\
\langle U \bar{\eta} \rangle &=& ev\, \langle z | 
 {i\over{(p^{2})^{2}}} \, 1_{-}1_{+} |z'\rangle.
\end{eqnarray} 
In actual calculations one handles $\theta$-diagonal elements, which
can be read off by using
$\langle \theta, \bar{\theta}|1_{+}1_{-}|\theta, \bar{\theta}\rangle =1$
and $\langle \theta, \bar{\theta}|Y|\theta, \bar{\theta}\rangle =2$.

In Eq.~(\ref{Gammaone}) we construct the one-loop effective action.
There we retain only terms with no $D_{\alpha}$ and
$\bar{D}_{\dot{\alpha}}$ acting on $(\Phi_{\rm c},V_{\rm c})$.
Let us first consider the contribution from the $\eta
\bar{\eta}$ loop, which is best calculated by first differentiating it
with respect to $V_{\rm c}$. Evaluating $2e\int d^{8}z\, \Xi\,
\langle \eta (z)\bar{\eta}(z)\rangle^{(0)}$ using
Eq.~(\ref{etapropagator}) and functionally-integrating back 
over $V_{\rm c}$ then yields, in our approximation,  
\begin{equation}
2e \langle x|  i\, e^{\tau p^{2}}/ p^{2}|x\rangle\, V_{\rm c}. 
\end{equation} 
This leads to  the UV divergent term in Eq.~(\ref{Gammaone})

The effective action coming from the $U$ loop is written as
$(i/2)\int d^{8}z \langle z|\ln M' |z \rangle$.
In the present approximation, 
$M' =-2 [\, p^{2}+ (\mu^{2}/p^{2})Y]$ with
$\mu^{2}= 2e^{2}\bar{\Phi}_{\rm c}\Xi \Phi_{\rm c}$.
One can calculate $(i/2) \langle z|\ln  M'|z\rangle$ by first
differentiating with respect to $\mu^{2}$, with the result 
\begin{equation}
\langle x| i/\{p^{2}(p^{2} -\mu^{2})\}|x\rangle.
\end{equation} 
This is convergent in (1+2) dimensions 
and, upon integrating over $\mu^{2}$, yields
$- |\mu|/(2\pi)$, which gives the second term in Eq.~(\ref{Gammaone}).


\section{gauge fixing}
In this appendix we show, for completeness, that the supertrace
identity~(\ref{DVcl}) is promoted to a background gauge invariant
form via gauge fixing. 
The gauge-fixing term $S^{\rm gf} =\int d^{8}z\,  (-{1\over{8 }})\,
(\bar{D}^{2}U)D^{2}U$ leads to an additional
contribution to the supercurrent ${\cal R}_{\alpha\dot{\alpha}}$ of
Eq.~(\ref{SCclass}), 
${\cal R}_{\alpha\dot{\alpha}}^{\rm gf} = (1/24)
[-(D_{\alpha}\bar{D}^{2}U)\bar{D}_{\dot{\alpha}}D^{2}U + \cdots]$; 
see Ref~[\onlinecite{CPS}] for the explicit form which is rather involved.
Its supertrace is written as
$\bar{D}^{\dot{\alpha}}{\cal R}^{\rm gf}_{\alpha \dot{\alpha}}  
= {\cal O}^{\rm gf}_{\alpha}
+ {\cal A}^{\rm gf}_{\alpha}$ with
\begin{eqnarray}
{\cal O}^{\rm gf}_{\alpha} &=& -(1/48)D_{\alpha}
\bar{D}^{2}(U[D^{2},\bar{D}^{2}]U),   \nonumber \\
{\cal A}^{\rm gf}_{\alpha}  
&=& 2 W_{\alpha}^{U} \delta S^{\rm gf}/\delta U
-  (D_{\alpha} U)p^{2}\bar{D}^{2}U,
\end{eqnarray}
where $W_{\alpha}^{U}= -{1\over{4}}\bar{D}^{2}D_{\alpha}U$.
Here ${\cal O}^{\rm gf}_{\alpha}$ is an explicit conformal breaking term
(which, on replacing $D_{\alpha}\bar{D}^{2}\rightarrow
\bar{D}^{2}D_{\alpha}$, may partly be absorbed into 
${\cal R}^{\rm gf}_{\alpha \dot{\alpha}}$). 
For the full supercurrent 
${\cal R}^{\rm full}_{\alpha\dot{\alpha}}
={\cal R}_{\alpha\dot{\alpha}}
+ {\cal R}^{\rm gf}_{\alpha\dot{\alpha}}$ the anomaly 
${\cal A}^{\rm full}_{\alpha} = {\cal A}_{\alpha} 
+ {\cal A}^{\rm gf}_{\alpha}$ 
is cast in a neat form
\begin{equation}
{\cal A}^{\rm full}_{\alpha} 
=  2 (W_{\rm c})_{\alpha} {\delta S'\over{\delta V_{\rm c}}} 
+ 2 W_{\alpha}^{U} {\delta S'\over{\delta U}} 
+ {1\over{2}} (D_{\alpha} U)\bar{D}^{2}
{\delta S'\over{\delta U}}+ \cdots, 
\end{equation}
where the omitted terms are the same as those in Eq.~(\ref{Aalpha}), 
except that $\nabla_{\alpha}[V]$ there is replaced by 
the background covariant derivative $\nabla_{\alpha}[V_{\rm c}]$; 
$S' = S[V] + S^{\rm gf}[U]$ stands for the full action. 
We note that ${\cal A}^{\rm full}_{\alpha}$ as well as the
supercurrent ${\cal R}^{\rm full}_{\alpha\dot{\alpha}}$ preserves
background gauge invariance and parity, which is crucial 
for determining its quantum form in Sec. IV.


\section{anomaly}
In this appendix we verify the discussion of Sec.~IV
by a direct one-loop calculation of the  anomaly ${\cal A}_{\alpha}$ in
Eq.~(\ref{Aalpha}).   As noted in Sec.~IV, one may set 
$\bar{\Phi}_{\rm c}, \Phi_{\rm c} \rightarrow 0$ in calculating 
${\cal A}_{\alpha}$.
The $U$ sector in the one-loop action $S_{\rm q}$ thereby becomes a free
field and does not contribute to the quantum anomaly at one loop. 
Thus $W_{\alpha} (\delta S/\delta V) =0$ in Eq.~(\ref{Aalpha}).
Similarly,  we conclude that $G (\delta S/ \delta V) =0$ 
in Eq.~(\ref{Oprime}).

The possible one-loop anomaly derives from the matter 
$(\eta, \bar{\eta})$ sector alone. 
Let us first consider 
$(\nabla_{\alpha}\Phi) (\delta S/\delta \Phi)$, which, using
Eq.~(\ref{etapropagator}), is cast in the form 
\begin{equation}
i\langle z| D_{\alpha} e^{\tau {\cal D}} 1_{-}|z \rangle
\approx 
-i  (W_{\rm c})_{\alpha}\tau \langle x| e^{\tau p^{2}} |x\rangle,
\end{equation}
for $\tau\rightarrow 0_{+}$, where one reaches the right-hand side 
by expanding the heat kernel in powers of
$W^{\alpha}_{\rm c}\nabla_{\alpha}$ and on taking the
$\theta$-diagonal matrix element. 
This leads to Eq.~(\ref{Danomaly}).
Similarly, one can verify $\Phi (\delta S/\delta \Phi) = 0$ by a direct
calculation of $i\langle z| e^{\tau {\cal D}} 1_{-}|z\rangle$.



\begin{thebibliography}{99}

\bibitem{B}  E. B. Bogomol'nyi, Sov. J. Nucl. Phys.  {\bf 24}, 449
(1976).

\bibitem{WO} 
E. Witten and D. Olive, Phys. Lett.  {\bf 78B}, 97 (1978).


\bibitem{W} E. Witten, Nucl. Phys.  {\bf B202}, 253 (1982).


\bibitem{Schon}  J. F. Schonfeld, Nucl. Phys. {\bf B161}, 125  (1979).


\bibitem{KR} R. K. Kaul and R. Rajaraman, 
Phys. Lett. {\bf 131B}, 357  (1983).


\bibitem{H} H. Yamagishi, Phys. Lett.  {\bf 147B}, 425  (1984).


\bibitem{IM} C. Imbimbo and S. Mukhi, Nucl. Phys. {\bf B247}, 471 (1984).


\bibitem{CM} A. K. Chatterjee and P. Majundar,  
Phys. Lett. {\bf 159B}, 37  (1985).


\bibitem{U} A. Uchiyama, Progr. Theor. Phys. {\bf 75}, 1214 (1986).


\bibitem{RN}   A. Rebhan and  P. van Nieuwenhuizen, 
Nucl. Phys. {\bf B508}, 449 (1997).


\bibitem{NSNR}  H. Nastase, M. Stephanov, P. van Nieuwenhuizen, 
and A. Rebhan, Nucl. Phys. {\bf B542}, 471 (1999).


\bibitem{GJ}  N. Graham and R. L. Jaffe, Nucl. Phys. {\bf B544}, 432 
(1999).

\bibitem{SVV} M. Shifman, A. Vainshtein, and M. Voloshin, 
Phys. Rev. D {\bf 59}, 045016 (2000).

\bibitem{LSV} 
A. Losev, M. Shifman, and A. Vainshtein, 
New J. Phys. {\bf 4}, 21 (2002).


\bibitem{RNW} A. Rebhan, P. van Nieuwenhuizen, and R. Wimmer,
New J. Phys. {\bf 4}, 31 (2002).


\bibitem{GRNW} A. S. Goldhaber, A. Rebhan, P. van Nieuwenhuizen, 
and R. Wimmer, Phys. Rep. {\bf 398}, 179 (2004).


\bibitem{SS} A. Salam and J. Strathdee,  Nucl. Phys. {\bf B76}, 477
(1974);
S. Ferrara, J. Wess, and B. Zumino, Phys. Lett. {\bf 51B}, 239 (1974).

\bibitem{WB} J. Wess and J. Bagger, {\sl Supersymmetry and supergravity},
(Princeton University Press, Princeton, 1992).

\bibitem{GSR}
M.T. Grisaru, W. Siegel, and M. Rocek, 
Nucl. Phys. {\bf B159}, 429 (1979). 


\bibitem{FN}  K. Fujikawa and P. van Nieuwenhuizen, 
Ann. Phys. (N.Y.) {\bf 308}, 78 (2003);

K. Fujikawa, A. Rebhan, and P. van Nieuwenhuizen, 
Int. J. Mod. Phys. A {\bf 18}, 5637 (2003).


\bibitem{KScc}  K. Shizuya, Phys. Rev. D {\bf 69}, 065021 (2004);
 Phys. Rev. D {\bf 70}, 065003 (2004).


\bibitem{FZ} S. Ferrara and B. Zumino, Nucl. Phys. {\bf B87}, 207 (1975).


\bibitem{CPS}
T. E. Clark, O. Piguet, and K. Sibold, Nucl. Phys.  {\bf B143},
445 (1978). 


\bibitem{Schm} 
J. R. Schmidt,  Phys. Rev. D {\bf 46}, 1839 (1992).


\bibitem{LM} 
B.-H. Lee and H. Min, Phys. Rev. D {\bf 51}, 4458 (1995).


\bibitem{Vass}  D. V. Vassilevich, Phys. Rev. D {\bf 68}, 045005 (2003).


\bibitem{RNWvor}  A. Rebhan, P. van Nieuwenhuizen, 
and R. Wimmer, Nucl. Phys.  {\bf B679}, 382 (2004). 


\bibitem{Fayet}  P. Fayet, Il Nuovo Cimento {\bf 31A}, 626 (1976).


\bibitem{FIL}  P. Fayet and J. Iliopoulos, Phys. Lett. {\bf 51 B}, 461
(1974).

\bibitem{WZQED} J. Wess and B. Zumino, Nucl. Phys.  {\bf B78}, 1 (1974).


\bibitem{DDT} S. C. Davis, A.-C. Davis, and M. Trodden,
Phys.  Lett. {\bf B 405}, 257 (1997).


\bibitem{WL} W. Lang, Nucl. Phys.  {\bf B150}, 201 (1979). 


\bibitem{CPStwo}
T. E. Clark, O. Piguet, and K. Sibold, Nucl. Phys.  {\bf B172},
201 (1980). 


\bibitem{KS}  K. Shizuya, Phys. Rev. D {\bf 35}, 1848 (1987). 
For related studies of effective actions and chiral anomalies in
superspace, see also,  
K. Shizuya and Y. Yasui, Phys. Rev. D {\bf 29}, 1160 (1984);  
K. Konishi and K. Shizuya, Nuovo Cimento {\bf 90A}, 111 (1985).


\bibitem{CCJ} C. G. Callan, S. Coleman, and R. Jackiw, 
Ann. Phys. (N.Y.) {\bf 59}, 42 (1970);

R. Jackiw, in {\sl Lectures on Current Algebra and Its
Applications}, (Princeton University Press, Princeton, 1972).


\bibitem{JT} S. Deser, R. Jackiw, and S. Templeton,
Ann. Phys. (N.Y.) {\bf 140} 372 (1982).  


\bibitem{F}  K. Fujikawa, Phys. Rev. Lett. {\bf 42}, 1195 (1979);
Phys. Rev. D {\bf 21}, 2848 (1980).


\bibitem{Ftwo}  K. Fujikawa, Phys. Rev. Lett. {\bf 44}, 1733 (1980);
Phys. Rev. D {\bf 23}, 2262 (1981).


\bibitem{CW}  S. Coleman and E. Weinberg, 
Phys. Rev. D {\bf 7}, 1888 (1973);

R. Jackiw, Phys. Rev. D {\bf 9}, 1686 (1974).


\bibitem{FNPRS} 
W.~Fischler, H.~P.~Nilles, J.~Polchinski, S.~Raby, and 
L.~Susskind, Phys. Rev. Lett. {\bf 47}, 757 (1981).


\end{thebibliography}
\end{document}